\documentclass{article}
\usepackage{amsmath}
\usepackage{graphicx}

\begin{document}

\title{Dark Hole Maintenance and A Posteriori Intensity Estimation in the Presence of Speckle Drift in a High-Contrast Space Coronagraph}

\maketitle

\author{Leonid Pogorelyuk and N. Jeremy Kasdin}

\begin{abstract}

Direct exoplanet imaging via coronagraphy requires maintenance of high
contrast in a dark hole for lengthy integration periods. Wavefront
errors that change slowly over that time accumulate and cause systematic
errors in the star's Point Spread Function (PSF) which limit the achievable signal-to-noise ratio of the planet. In this paper we show that estimating
the  speckle drift can be achieved via intensity measurements
in the dark hole together with dithering of the deformable mirrors
to increase phase diversity. A scheme based on an Extended Kalman Filter
and Electric Field Conjugation is proposed for maintaining the dark
hole during the integration phase. For the post-processing phase,
an {\it a posteriori} approach is proposed to estimate the realization of
the PSF drift process and the intensity of the planet light incoherent with
the speckles.

\end{abstract}

\section{Introduction} \label{sec:intro}

Direct imaging of Earthlike exoplanets requires detecting light from a planet
whose intensity is $10^{-10}$ times dimmer than the light of its host star.  Doing so necessitates the control of diffraction in the telescope to remove the residual light from the stellar PSF at the planet location.  Over the last several decades, several families of approaches to control diffraction at this level, collectively referred to as coronagraphy, have been proposed and implemented in the lab and on ground telescopes.  However, achieving the extremely high contrast needed to image and characterize rocky planets like Earth requires extreme control of the wavefront in the coronagraph.  This is accomplished by going to space, where atmospheric interference is eliminated and thermal and dynamic instabilities are minimized, and by implementing active control of the wavefront via deformable mirrors to correct for the slowly changing distortions within the telescope. In fact, the first fully capable in-space high-contrast coronagraph with wavefront control is planned for NASA's next large observatory, the 
Wide-Field Infra-Red Survey Telescope (WFIRST) Coronagraph Instrument
(CGI) (\cite{demers2015requirements,tang2015wfirst}).  This instrument will demonstrate the critical technologies and algorithms needed for future missions.

Wavefront errors in a coronagraph are typically divided into Low-Order and High-Order modes.
The control of low spatial frequency and high temporal frequency wavefront errors
(LOWFE) has been successfully demonstrated in the laboratory via Low Order
Wavefront Sensing and Control (LOWFS/C) (\cite{shi2015low,shi2017dynamic}) that
employs Fast Steering Mirrors (FSM) (\cite{patterson2015control}). However,
these methods address just the first few spatial modes of the wavefront
phase at the pupil plane (in terms of Zernike polynomials). High-order wavefront error (HOWFE) caused, for example, by inaccuracies in coronagraph masks and high-frequency surface errors,
are addressed via Focal Plane Wavefront Correction (FPWC) methods
such as Electric Field Conjugation (EFC) (\cite{give2007electric,give2007broadband})
and Stroke Minimization (\cite{pueyo2009optimal}). These methods employ
the deformable mirrors (DMs) both to correct the wavefront and to increase phase diversity in order to estimate the residual star light
electric field (speckles) from intensity measurements in the high-contrast
region (dark hole) of the science camera plane (\cite{paxman1992joint})  This avoids the introduction of uncorrectible non-common path errors introduced by a separate sensing optical path, as in ground adaptive optics.

FPWC is designed to create the dark hole by converging to the appropriate
DM control settings, which are then kept constant throughout the long
integration phase as the telescope is pointed at the target star.  
Nevertheless, even with an optimal wavefront control system and perfectly stable telescope, these control techniques typically are only capable of reducing the stellar speckle intensity to the level of the planet; sophisticated image analysis approaches are also required to separate the incoherent planet light from the residual stellar halo.
The current approach for extracting a planet from the background limited image is to subtract an estimated reference Point Spread Function from the composite image after taking a series of long exposures while pointed at the target start.  This has been commonly referred to as ``post-processing'' or ``PSF subtraction.''
The resulting intensity estimate is presumed to contain the planet signal, other sources of light incoherent with the star (e.g. zodi, dark current, etc.), and starlight residuals from systematic errors in the PSF. A key goal of post-processing approaches is to remove these systematic errors.  Approaches to mitigating them currently in use on the ground or proposed for space include rotating the field of view (Angular Differential Imaging or ADI) (\cite{lowrance1998coronagraphic,marois2006angular}),
projecting the data onto precomputed PSF subspaces (\cite{soummer2012detection}),
or using bright stars to obtain reference PSFs (\cite{mawet2011dim}).

All of the approaches to post-processing assume the residual stellar PSF after wavefront control does not change during the final integration (implying fixed DM settings and no observatory instabilities). In reality, the PSF will change due to thermal and dynamic drift of the telescope as well as DM drift, translating directly into errors in estimates of planet intensity and limits on the achievable Signal-to-Noise Ratio (SNR) (\cite{nemati2017sensitivity}). As a result, the potential of high-order disturbances such as mechanical and thermal stresses over periods of tens of days can impose tight stability requirements on the optical elements of the observatory and instrument (\cite{shaklan2011stability,stahl2013engineering}).
One alternative that relaxes stability requirements is the recently proposed Linear Dark Field Control (LDFC). LDFC keeps actuating
the DM to maintain the high contrast by utilizing measurements of
speckles outside the dark hole (\cite{miller2017spatial,guyon2017spectral}).
LDFC is the first closed loop approach to take into account wavefront error drift, although it relies on both measuring speckles outside of the dark
hole and on a high-fidelity model relating them to the speckles inside the dark hole.
In this paper, we derive a post-processing {\it a posteriori} intensity estimator
which takes into account speckle drift and the history of DM actuations.
This algorithm can then be used in conjunction with any closed loop dark-hole maintenance scheme. One such scheme, which relies solely on intensity
measurements in the dark hole, is also proposed here.

Our key finding is that small DM actuations (dither) during the integration phase are necessary for both post-processing and online PSF estimation for closed loop control. Although it has been previously suggested that small, unknown perturbations of the speckles might reduce their spatial variability and thus increase the planet SNR (\cite{angel1994ground}), it was later shown not to be the case in practice (\cite{sivaramakrishnan2002speckle}). Consequently, one must incorporate additional information from the influence of DM dithering on the speckles, as shown in this paper.

Section \ref{sec:overview} is an overview of the most common current focal-plane wavefront control approaches, Electric Field Conjugation (EFC), and
PSF subtraction for planet detection. In section \ref{sec:a_posterior_estimation}, we introduce
the post-processing (or offline) intensity estimator. In section \ref{sec:closed_loop}, we propose a closed loop control and estimation scheme for maintaining the contrast in the dark hole and show that it benefits from small random DM actuations (dither) to increase phase diversity. Finally, section \ref{sec:numerical_results} presents a numerical study of the proposed algorithms and their combined potential for estimating planet intensity in the presence of instabilities, thus potentially relaxing the severe telescope stability requirements.

\section{\label{sec:overview}Creating the Dark Hole and PSF subtraction}

\subsection{\label{sub:EFC}Dark Hole Creation via EFC}

\begin{figure}
\includegraphics[width=\textwidth]{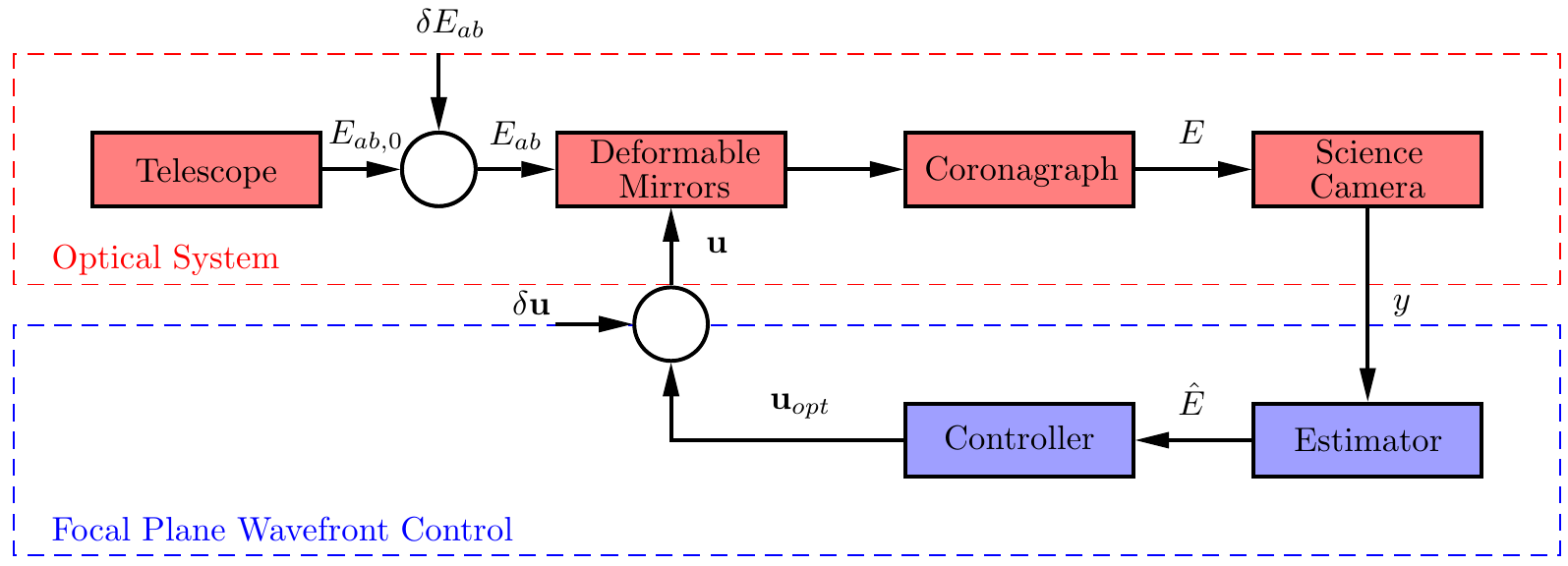}
\caption{\label{fig:diagram}The aberrated electric field after low-order correction, $E_{ab}$,
is slightly modified by the DM before passing through the coronagraph to produce the final field, $E$, whose intensity is
detected at the science camera as a photon count, $y$ (which grows linearly with intensity, $I=\left|E\right|^{2}$).
FPWC estimates the electric field at the camera (denoted by \textasciicircum{}), and computes the optimal
control, $\mathbf{u}_{opt}$, to minimize speckle intensity. Probes or dither
($\delta \mathbf{u}$) are added to the optimal actuations, $\mathbf{u}_{opt}$, in
order to increase phase diversity. The drift of the aberrated electric
field, $\delta E_{ab}$, is unknown and the estimation of its effects
on the focal plane electric field, $E$, is the subject of section \ref{sec:a_posterior_estimation}.}
\end{figure}

Creating a high-contrast dark hole to reduce the background due to the stellar halo minimizes the shot noise due to
speckles and thus increases the post-processing SNR. It has long been known that a coronagraph alone is not sufficient to create the needed high contrast due to wavefront error in the telescope.  Thus, wavefront control is introduced to correct the distortions and recover the needed contrast.
This is done iteratively
by estimating the electric field of the speckles at the focal plane,
$E$, and applying some optimal control (in a sense which will
be discussed later), $\mathbf{u}_{opt}$, to reduce the resulting field intensity
$I=\left|E\right|^{2}$; see Fig.~\ref{fig:diagram}.

Using only Fourier optics, it can be shown that the electric field at some plane after the first DM, $E_{p,1}$, is
a linear operator, ${\cal C}_{1}$, acting on the wavefront leaving the deformable mirrors,
\begin{equation}
E_{p,1}={\cal C}_{1}\left\{ E_{ab}e^{i\phi\left(\mathbf{u}_{1}\right)}\right\} ,
\end{equation}
where $E_{ab}$ is the incoming, aberrated wavefront and 
$\phi\left(\mathbf{u}_{1}\right)$ is the phase induced by the command $\mathbf{u}_{1}$ of a first DM. Similarly, the field at some plane after the second DM is given by a second operator, ${\cal C}_{2}$,
\begin{equation}
E_{p,2}={\cal C}_{2}\left\{ E_{p,1}e^{i\phi\left(\mathbf{u}_{2}\right)}\right\} ,
\end{equation}
etc.\ until the focal plane of the detector which we define with a composite operator, ${\cal C}$,
\begin{equation}
E={\cal C}\left\{ E_{ab}, \mathbf{u}_{1},\mathbf{u}_{2}, ...\right\}={\cal C}\left\{ E_{ab}, \mathbf{u}\right\}.
\end{equation}
It has been shown (\cite{give2007broadband}) that multiple deformable mirrors are required to correct for both phase and amplitude disturbances in the aberrated field, $E_{ab}$.

Figure \ref{fig:diagram} illustrates a general closed loop control approach with some ``optimal'' control setting, $\mathbf{u}_{opt}$, computed
at iteration $k$. Throughout the paper, $E_{i,j}(k)$ denotes the electric field $E$ at pixel $i,j$ and time $k$ and the vector $\mathbf{E}(k)$ denotes the real and imaginary parts of all $E_{i,j}(k)$.
The effect of small deviations from $\mathbf{u}_{opt}(k)$
on the focal plane electric field, $\mathbf{E}(k)$, can be approximated to first order
by,
\begin{eqnarray}
\mathbf{E}(k)  & = \mathbf{E} \left( E_{ab}(k), \mathbf{u}_{opt}(k) + \delta \mathbf{u}(k) \right) \approx
\\
& \approx  \mathbf{E} \left( E_{ab}(k), \mathbf{u}_{opt}(k) \right) + {\cal G}\delta \mathbf{u}(k),\label{eq:linearity_assumption}
\end{eqnarray}
where ${\cal G}$ is the Jacobian matrix of the coronagraph operator, which is either precomputed or estimated in real time (\cite{sun2018identification}),
and $\delta \mathbf{u}(k)$ are small actuations.  The small perturbations to the DM setting are either predetermined probes used to estimate the field for creating the dark hole (\cite{give2011pair}) or randomly generated to produce dither, as described below.
In both cases, these small actuations introduce the phase diversity necessary for estimating the electric field at the science camera (\cite{groff2015methods}).

Given the $k$th estimate of the field $\hat{\mathbf{E}}(k)$ (with
$\delta \mathbf{u}=0$), the control at the next step can be computed by various
methods such as Speckle Nulling (\cite{borde2006high}), Electric Field
Conjugation (EFC) (\cite{give2007electric}) and/or Stroke Minimization (\cite{pueyo2009optimal}).
Specifically, for EFC, the optimal control minimizes a weighted sum
of the field intensity and control energy at step $k$. The resulting command is given
by,

\begin{equation}
\mathbf{u}_{opt}(k+1)=\mathbf{u}_{opt}(k)-\left({\cal G}^{t}{\cal G}+\alpha{\cal I}\right)^{-1}{\cal G}^{t}\hat{\mathbf{E}}(k),\label{eq:EFC}
\end{equation}
where $\alpha$ is some constant, ${\cal I}$ is the identity matrix and $\cdot^{t}$ stands for the matrix transpose.
Note that EFC treats the open loop electric field as constant ($E_{ab}(k)=E_{ab}(0)$),
which results in systematic errors in the estimate of the planet intensity
when speckle drift is present. This issue is addressed
next.

\subsection{PSF Subtraction}

The current approach to processing for planet detection is to take long exposure images  (or many stacked exposures) once a dark hole has been established, creating the final science image.
PSF subtraction based methods are then employed to remove the residual speckles remaining after control and to estimate the planet intensity (e.g., roll subtraction (\cite{lowrance1998coronagraphic}),
ADI (\cite{marois2006angular}), or KLIP (\cite{soummer2012detection})). All of these techniques assume
that the speckle field does not change in time ($E(k)=E(0),\:\forall k$) during the final integration step.
These approaches produce an estimate of the intensity of all light incoherent with the
speckles via
\begin{equation}
\hat{I}_{ij}=\frac{1}{T}\underset{k=1}{\overset{T}{\sum}}\left(\beta^{-1}y_{ij}(k)-\left|\hat{E}_{ij}(0)\right|^{2}\right),\label{eq:PSF_sub}
\end{equation}
where, $i,j$ are indices of a single detector pixel, $T$ is the
total number of exposure frames, $\beta$ is the mean number of photons
per unit intensity expected to arrive at a single pixel in the duration
of a single frame and $y$ is the number of photons that were actually
detected. Initial errors in $\left|\hat{E}_{ij}(0)\right|^{2}$, mechanical
stresses during pointing and roll maneuvers, and thermal instabilities
of the optical system all contribute to the systematic error directly affecting the estimate $\hat{I}_{ij}$.

Additional sources contributing to the number of ``detected'' photons (e.g.
zodiacal light and dark current), have low spatial frequencies and are
constant in time. We assume that they produce electrons which are
Poisson distributed and therefore indistinguishable from the planet
induced electrons. 
Consequently, we lump all of these sources together into one term, 
\begin{equation}
I_{ij}=I_{ij}^{P}+I_{ij}^{Z}+I_{ij}^{D},
\end{equation}
where $I_{ij}^{P}$ and $I_{ij}^{Z}$ are the intensities of the light from
the planet and zodi respectively and $I_{ij}^{D}$ is the effective intensity
of the dark current (i.e., an intensity of light that, given the efficiency
of the detector, would produce electrons distributed identically to
the thermal electrons).

To simplify matters further, we will ignore $I_{ij}^{Z}$ 
and scale all physical quantities
such that $\beta=1$. Since the number of detected photons per pixel, $y_{ij}$,
is Poisson distributed,

\begin{equation}
y_{ij}\sim\mathrm{Poiss}\left(I_{ij}+\left|E_{ij}\right|^{2}\right),\label{eq:photons_count}
\end{equation}
the signal-to-noise ratio (SNR) after PSF subtraction is given by (\cite{nemati2017sensitivity})

\begin{equation}
\mathrm{SNR}_{ij}\left(T\right)=\frac{I_{ij}^{P}T}{\sqrt{\left(I_{ij}+\left|E_{ij}\right|^{2}\right)T+\left(\Delta\left|E_{ij}\right|^{2}\right)^{2}T^{2}}},\label{eq:SNR}
\end{equation}
where $\Delta\left|E_{ij}\right|^{2}$ is the systematic error in speckle
intensity. Note that this error limits the maximum attainable SNR
to $\frac{I_{ij}^{P}}{\Delta\left|E_{ij}\right|^{2}}$, indicating its strong dependence on reference PSF  errors and instabilities. More advanced methods which use multiple reference PSFs, reduce this limit (assuming the systematic errors associated with them are uncorrelated), but do not eliminate it completely.

In the remainder of this paper we relax the assumption that the residual speckle field is constant and allow for non-fixed DM actuation throughout the integration
phase to alleviate this issue of limited achievable SNR. To this end, we introduce an estimator for both the intensity of the incoherent light,
$\hat{I}_{ij}$, and the history of the speckle field, $\left\{\hat{E}_{ij}(k) \right\}_{k=0}^{T}$.

\section{\label{sec:a_posterior_estimation}A Posteriori Intensity Estimation}
In this section we describe a post-processing method for estimating the incoherent intensity, $I_{ij}$, given the history of measurements $y_{ij}$ and controls $\mathbf{u}$. It is independent of the algorithm used for choosing the controls in real-time, if any (one such algorithm is provided in Sec.~\ref{sec:closed_loop}).
The underlying assumption throughout this section is that we begin with a dark hole created by some optimal control process such as EFC (see Sec.~\ref{sub:EFC}), and that all changes in the speckle field and DM actuations are small deviations from their nominal values established at the end of that dark-hole generation process.

\subsection{\label{sub:drift_process}Speckle Drift Process}

Statistical methods exploiting phase diversity to estimate the electric field of the speckles, with or without DM
actuation, have been present in the literature for a long time (for a review, see \cite{rousset1999wave}). 
None of them, however, take into account the evolution of the speckle
field, which is a stochastic process with some initial distribution (of $E_{ij}(0)$).
Given $N$ pixels and $T$ frames, the joint distribution of $E_{ij}(k)$
consists of at least $N\cdot T$ \emph{dependent} random variables,
which renders the estimation problem computationally intractable.
We therefore introduce a simplifying assumption to be used for simulation
and estimation purposes in the remainder of the paper.

First, we slightly modify equation (\ref{eq:linearity_assumption}) to fit in a more general context than just dark hole creation,
\begin{equation}
E_{ij}\left(\mathbf{u},k\right)\approx E_{ij}^{OL}\left(E_{ab}(k),\mathbf{u}_{0}\right)+G_{ij}\Delta\mathbf{u}(k),\label{eq:EOL_def}
\end{equation}
and isolate the open loop behavior, $E^{OL}_{ij}(k)=E_{ij}^{OL}\left(E_{ab}(k),\mathbf{u}_{0}\right)$, of the speckles after the completion of the dark hole.  Here, $\mathbf{u}_0$ is the optimal dark hole control setting chosen to flatten the open loop field $E_{ij}^{OL}(0)$ and $\Delta \mathbf{u}(k)$ is a possible additional control setting on top of $\mathbf{u}_0$ to maintain the dark hole (see Sec.~\ref{sec:closed_loop}).
Second, we introduce the
assumption that the increments of the open loop speckle field $E_{ij}^{OL}(k)-E_{ij}^{OL}(k-1)$
are spatially and temporally independent (with respect to $i,j$ and
$k$), thus forming a Brownian random walk. The residual speckle field at each pixel can then be considered
individually and the probability density of its sample path written
as the product,

\begin{equation}
p\left(E_{ij}^{OL}(0),\cdots,E_{ij}^{OL}(T)\right)=p\left(E_{ij}^{OL}(0)\right)\underset{k=1}{\overset{T}{\prod}}p\left(E_{ij}^{OL}(k)-E_{ij}^{OL}(k-1)\right).
\end{equation}

Although the above model discards spatial and temporal correlations, it is a good starting point for pixel based estimation of the incoherent light. We are currently developing a reduced-order approach that takes spatial coupling into account; it will be presented in a future paper. 

This drift model is simplified further by assuming
\begin{eqnarray}
E_{ij}^{OL}\left(k\right)-E_{ij}^{OL}\left(k-1\right) & \sim{\cal N}\left(\mu_{ij},\Sigma_{ij}\right),\label{eq:field_increment}\\
E_{ij}^{OL}\left(0\right) & \sim{\cal N}\left(\mu_{ij,0},\Sigma_{ij,0}\right)\label{eq:field_init}
\end{eqnarray}
where, with a slight abuse of notation, the $E_{ij}$ are treated as
2D vectors consisting of the real and imaginary parts of the electric
field at pixel $i,j$ and ${\cal N}\left(\mu_{ij},\Sigma_{ij}\right)$ stands for the Normal distribution
with mean $\mu_{ij}$ and covariance matrix $\Sigma_{ij}$. In a simulation
setting (see section \ref{sec:numerical_results}), the parameters $\mu_{ij}$ and $\Sigma_{ij}$ can be estimated via a Monte-Carlo method. 

\subsection{\label{sub:estimator}Speckle Drift Process Estimator}

Equipped with a prior distribution for the field at pixel $i,j$, it becomes possible to define 
corresponding estimators of the incoherent intensity, $\hat{I}_{ij}$, and the posterior field, $\hat{E}_{ij}^{OL}(k)$,
based on the observations $y_{ij}(k)$. Although any
such estimators would be sub-optimal as they discard spatial correlation
between pixels, they allow for a computationally tractable solution. Before defining a simultaneous estimate for both the intensity and the field history, we consider each of their estimates given the other.

When the field sample path, $\left\{ E_{ij}^{OL}(k)\right\}=\left\{E_{ij}^{OL}(0),\cdots,E_{ij}^{OL}(T)\right\}$,
is given, Eq.~(\ref{eq:PSF_sub}) can be extended to a sample mean estimator,

\begin{equation}
\hat{I}_{ij}=\frac{1}{T}\underset{k=1}{\overset{T}{\sum}}\left(y_{ij}(k)-\left|E_{ij}^{OL}(k)+G_{ij}\Delta\mathbf{u}(k)\right|^{2}\right),\label{eq:intermediate_estimator}
\end{equation}
where $\Delta\mathbf{u}(k)=\mathbf{u}(k)-\mathbf{u}_{0}$ are known DM controls (which could have possibly been chosen by the algorithm in Sec.~\ref{sec:closed_loop} or set to zero in the case of no control during the science observation).

Alternatively, given the incoherent intensity, $I_{ij}$, and the number
of photons observed, $\left\{{y}_{ij}(k)\right\}=\left\{y_{ij}\left(1\right),\cdots,y_{ij}\left(T\right)\right\}$,
a maximum {\it a posteriori} estimator for $\left\{ \hat{E}_{ij}^{OL}(k)\right\}=\left\{\hat{E}_{ij}^{OL}(0),\cdots,\hat{E}_{ij}^{OL}(T)\right\}$
is one which maximizes the conditional probability of $\left\{{y}_{ij}(k)\right\}$. In other words,

\begin{equation}
\left\{ \hat{E}_{ij}^{OL}(k)\right\}=\underset{\left\{E_{ij}^{OL}(k)\right\}}{\mathrm{argmax}}\:p\left(\left.\left\{{y}_{ij}(k)\right\}\right|I_{ij},\left\{E_{ij}^{OL}(k)\right\}\right)p\left(\left\{E_{ij}^{OL}(k)\right\}\right),
\end{equation}
where $p\left(\left\{E_{ij}^{OL}(k)\right\}\right)$ is the prior distribution of the electric field (obtained from Eqs. (\ref{eq:field_increment}), (\ref{eq:field_init})), and the conditional probability of $\left\{{y}_{ij}(k)\right\}$ given $I_{ij}$ and $\left\{E_{ij}^{OL}(k)\right\}$ is

\begin{equation}
p\left(\left.\left\{{y}_{ij}(k)\right\}\right|I_{ij},\left\{E_{ij}^{OL}(k)\right\}\right)=\underset{k=1}{\overset{T}{\prod}}\frac{\left(I_{ij}+\left|E_{ij}^{OL}(k)+G_{ij}\Delta\mathbf{u}(k)\right|^{2}\right)^{y_{ij}(k)}}{y_{ij}(k)!}e^{-\left(I_{ij}+\left|E_{ij}^{OL}(k)+G_{ij}\Delta\mathbf{u}(k)\right|^{2}\right)},\label{eq:observation_probability}
\end{equation}
since $\left\{{y}_{ij}(k)\right\}$ are independently Poisson distributed. Note that while the intensities of the speckles at different pixels and times are correlated, the {\it conditional} distributions of photons at a {\it single} pixel are uncorrelated in time.

We therefore introduce the following mixed estimator

\begin{eqnarray}
\tilde{I}_{ij}\left(\left\{E_{ij}^{OL}(k)\right\}\right) \equiv & \frac{1}{T}\underset{k=1}{\overset{T}{\sum}}\left(y_{ij}(k)-\left|E_{ij}^{OL}(k)+G_{ij}\Delta\mathbf{u}(k)\right|^{2}\right)\label{eq:estimator1}\\
\left\{\hat{E}_{ij}^{OL}(k)\right\} = & \underset{\left\{E_{ij}^{OL}(k)\right\}}{\mathrm{argmax}}\:p\left(\left.\left\{{y}_{ij}(k)\right\}\right|\tilde{I}_{ij}\left(\left\{E_{ij}^{OL}(k)\right\}\right),\left\{E_{ij}^{OL}(k)\right\}\right)p\left(\left\{E_{ij}^{OL}(k)\right\}\right)\label{eq:estimator2}\\
\hat{I}_{ij} = & \tilde{I}_{ij}\left(\left\{\hat{E}_{ij}^{OL}(k)\right\}\right)\label{eq:estimator3}
\end{eqnarray}
with $\tilde{I}_{ij}$ taking the role of an ``intermediate-step
sample mean estimator'' suggested by Eq.~(\ref{eq:intermediate_estimator}).

The estimator defined by Eqs. (\ref{eq:estimator1})-(\ref{eq:estimator3}) is not
unbiased or efficient due to the mixed treatment of the intensity
and the electric field. However, it only requires optimization over the
electric fields $E_{ij}^{OL}(0),\cdots,E_{ij}^{OL}(T)$, which have
the same order of magnitude and for which we can obtain a good initial
guess (see section \ref{sec:closed_loop}).

\subsection{\label{sub:estimation_algorithm}Offline Estimation Algorithm}

One should be cautious when applying Eq.~(\ref{eq:estimator1}), as the
intensity cannot possibly be lower than the intensity of the dark
current, $I_{ij}\ge I_{D}>0$, and negative values of $\tilde{I}_{ij}$
may give negative probabilities in (\ref{eq:estimator2}). For numerical
purposes we define
\begin{equation}
\tilde{I}_{ij}^{\#}=\begin{cases}
\tilde{I}_{ij} & \tilde{I}_{ij}>I_{D}\\
I_{D} & otherwise
\end{cases},\label{eq:constrained_I}
\end{equation}
and

\begin{equation}
\frac{d}{d\tilde{I}_{ij}}\tilde{I}_{ij}^{\#}=\begin{cases}
1 & \tilde{I}_{ij}>I_{D}\\
0 & otherwise
\end{cases}.
\end{equation}

Putting Eqs. (\ref{eq:field_increment}),(\ref{eq:field_init}),(\ref{eq:observation_probability})
and (\ref{eq:estimator1})-(\ref{eq:estimator3}) together, the cost to be minimized at each pixel is explicitly written as

\begin{eqnarray}
J_{ij}\left(\mathbf{E}_{ij}^{OL}\right) \equiv & -\log\left(p\left(\left.\mathbf{y}_{ij}\right|\tilde{I}_{ij}^{\#}\left(\mathbf{E}_{ij}^{OL}\right),\mathbf{E}_{ij}^{OL}\right)p\left(\mathbf{E}_{ij}^{OL}\right)\right)=\nonumber \\
 = & -\underset{k=1}{\overset{T}{\sum}}y_{ij}(k)\log\left(\tilde{I}_{ij}^{\#}\left(\mathbf{E}_{ij}^{OL}\right)+\left|E_{ij}^{OL}(k)+G_{ij}\Delta\mathbf{u}(k)\right|^{2}\right)+\nonumber \\
 + & \underset{k=1}{\overset{T}{\sum}}\left(\tilde{I}_{ij}^{\#}\left(\mathbf{E}_{ij}^{OL}\right)+\left|E_{ij}^{OL}(k)+G_{ij}\Delta\mathbf{u}(k)\right|^{2}\right)+\nonumber \\
 + & \underset{k=1}{\overset{T}{\sum}}\frac{1}{2}\left(\begin{bmatrix}\mathrm{Re}\left\{ E_{ij}^{OL}(k)-E_{ij}^{OL}(k-1)\right\} \\
\mathrm{Im}\left\{ E_{ij}^{OL}(k)-E_{ij}^{OL}(k-1)\right\} 
\end{bmatrix}-\mu_{ij}\right)^{t}\Sigma_{ij}^{-1}\cdot \nonumber \\
 & \cdot \left(\begin{bmatrix}\mathrm{Re}\left\{ E_{ij}^{OL}(k)-E_{ij}^{OL}(k-1)\right\} \\
\mathrm{Im}\left\{ E_{ij}^{OL}(k)-E_{ij}^{OL}(k-1)\right\} 
\end{bmatrix}-\mu_{ij}\right)+\nonumber \\
 + & \frac{1}{2}\left(\begin{bmatrix}\mathrm{Re}\left\{ E_{ij}^{OL}(0)\right\} \\
\mathrm{Im}\left\{ E_{ij}^{OL}(0)\right\} 
\end{bmatrix}-\mu_{ij,0}\right)^{t}\Sigma_{ij,0}^{-1}\cdot \nonumber \\
 & \cdot\left(\begin{bmatrix}\mathrm{Re}\left\{ E_{ij}^{OL}(0)\right\} \\
\mathrm{Im}\left\{ E_{ij}^{OL}(0)\right\} 
\end{bmatrix}-\mu_{ij,0}\right)+\underset{k=1}{\overset{T}{\sum}}\log 
\end{eqnarray}


This cost, $J_{ij}$, can be minimized with respect to $\left\{E_{ij}^{OL}(k)\right\}_{k=0}^{T}$
using standard gradient descent methods;  the final estimate of
the incoherent intensity is then found from

\begin{equation}
\hat{I}_{ij}=\tilde{I}_{ij}^{\#}\left(\underset{\left\{E_{ij}^{OL}(k)\right\}}{\mathrm{argmin}}\:J_{ij}\left(\left\{E_{ij}^{OL}(k)\right\}\right)\right).
\end{equation}

Note that the above offline estimate incorporates the closed loop
control $\Delta\mathbf{u}$ which has to be either predetermined or
chosen in real time. The numerical study in section \ref{sub:open_loop_simulation}
illustrates the benefits of non-zero predetermined $\Delta\mathbf{u}$,
while the next section discusses the choice of $\Delta\mathbf{u}$
based on an online estimator of the speckles.

\subsection{\label{sub:implementation}Implementation}
To summarize, the goal of the post-processing scheme is to estimate the constant incoherent signal, $I_{ij}$ (at pixel $i,j$), given the history of photon counts, $y_{ij}(0),\cdots,y_{ij}(T)$, and DM controls, $\Delta\mathbf{u}(0),\cdots,\Delta\mathbf{u}(T)$. This is done for each pixel separately, regardless of the scenario in which these measurements were obtained and the technique used to choose the controls.

The signal, $I_{ij}$, is estimated indirectly by first optimizing the cost function $J_{ij}$ with respect to the unknown history of the electric field at pixel $i,j$, denoted by $\hat{E}_{ij}^{OL}(0),\cdots,\hat{E}_{ij}^{OL}(T)$. The authors found the Dataflow Graphs approach utilized by TensorFlow (\cite{abadi2016tensorflow}), to be the most convenient for implementation purposes, since it reduces the problem to merely \emph{defining} the cost function in terms of the optimization parameters (or variables) and inputs. The optimization procedure itself can then be chosen among several standard gradient-decent algorithms and invoked in a straightforward manner (for the results in Sec.~\ref{sec:numerical_results}, the authors employed the TensorFlow implementation of the Adam Optimizer \cite{kingma2014adam}).

We therefore proceed by defining (in a hierarchical fashion) the cost function, $J_{ij}$, in terms of the inputs $\left\{y_{ij}(k)\right\}_{k=0}^{T}$, $\left\{\Delta\mathbf{u}(k)\right\}_{k=0}^{T}$ and the variables $\left\{\hat{E}_{ij}^{OL}(k)\right\}_{k=0}^{T}$: first, the intensity estimate is given by Eq.~(\ref{eq:estimator1}), then its constrained value is defined in Eq.~(\ref{eq:constrained_I}); finally, the cost function itself is given by Eq.~(\ref{eq:full_cost_func}).

Since $J_{ij}$ is a non-linear function, the initial guess of $\left\{\hat{E}_{ij}^{OL}(k)\right\}_{k=0}^{T}$ plays an important role in the accuracy of the final estimate. In the numerical simulations presented in Sec.~\ref{sec:numerical_results}, we obtained our initial guesses from the electric field estimator described in Sec.~\ref{sub:high_intensity_estimation} and Appendix \ref{sec:ekf_equations}, and used for control purposes during the observation scenario itself.

\section{\label{sec:closed_loop}Closed Loop Dark Hole Maintenance}

In this section we present a real-time feedback controller for maintaining a high contrast in the dark hole. It consists of a recursive estimator of the speckle field and a slightly modified EFC control law.
The estimation technique presented in the previous section could, in principle, be used instead of a recursive estimator, although it requires optimizing over the entire speckle history at every pixel, a computationally expensive task. As we will show, the
joint estimation and control problem is non-linear, and hence the
choice of optimal closed loop control, $\Delta\mathbf{u}$, is non-trivial.

\subsection{\label{sub:high_intensity_estimation}High-Intensity Regime Recursive
Estimation}

Assuming that the increments of the speckle field are spatially and
temporally independent and normally distributed (Eq.~(\ref{eq:field_increment})),
the corresponding state equation is

\begin{eqnarray}
\begin{bmatrix}\mathrm{Re}\left\{ E_{ij}^{OL}(k+1)\right\} \\
\mathrm{Im}\left\{ E_{ij}^{OL}(k+1)\right\} 
\end{bmatrix} & =\begin{bmatrix}\mathrm{Re}\left\{ E_{ij}^{OL}(k)\right\} \\
\mathrm{Im}\left\{ E_{ij}^{OL}(k)\right\} 
\end{bmatrix}+v_{E}(k)\label{eq:EKF_state1}\\
I_{ij}(k+1) & =I_{ij}(k)+v_{I}(k),\label{eq:EKF_state2}
\end{eqnarray}
where $v_{E}(k)\sim{\cal N}\left(\mu_{ij},\Sigma_{ij}\right)$ and
$v_{I}(k)\sim{\cal N}\left(0,\sigma_{I}^{2}\right)$ ($\sigma_{I}$
is non-zero for numerical purposes). Note that the control input $\mathbf{u}$
is excluded from the state equation contrary to its common formulation (\cite{groff2013kalman}).

To put the observation equation in a convenient form, we note that
the Poisson distribution of the number of photons, $y_{ij}(k)$, converges
to a Normal distribution, i.e.

\begin{equation}
y_{ij}(k)\sim{\cal N}\left(I_{ij}+\left|E_{ij}^{OL}(k)+G_{ij}\Delta\mathbf{u}(k)\right|^{2},I_{ij}+\left|E_{ij}^{OL}(k)+G_{ij}\Delta\mathbf{u}(k)\right|^{2}\right)
\end{equation}
for large values of the intensity. This gives an approximation of
the measurement equation,

\begin{equation}
y_{ij}(k)\approx I_{ij}+\left|E_{ij}^{OL}(k)+G_{ij}\Delta\mathbf{u}(k)\right|^{2}+\sqrt{I_{ij}+\left|E_{ij}^{OL}(k)+G_{ij}\Delta\mathbf{u}(k)\right|^{2}}w(k),\label{eq:EKF_observation}
\end{equation}
where $w(k)\sim{\cal N}\left(0,1\right)$. 

Equations (\ref{eq:EKF_state1}), (\ref{eq:EKF_state2}) and (\ref{eq:EKF_observation})
allow for a standard formulation of an Extended Kalman Filter (EKF) with non-additive noise (see Appendix \ref{sec:ekf_equations}). 
The approximation in Eq.~(\ref{eq:EKF_observation}) breaks down when the average number of photons per pixel per frame is significantly
less than one. In that case one may either increase the exposure time
for each frame or combine several frames together.

\subsection{\label{sub:EFC_dithering}EFC with Dithering}

Unfortunately, a filter combined with a naive choice of a control
law may result in a completely wrong estimate of the speckle field.
It can be shown that the optimal EFC control law based on
a recursive EKF estimate of the field $\hat{E}^{OL}(k)$,

\begin{equation}
\Delta\mathbf{u}_{opt}(k+1)=-\left({\cal G}^{t}{\cal G}+\alpha{\cal I}\right)^{-1}{\cal G}^{t} \hat{\mathbf{E}}^{OL}(k).\label{eq:opt_u}
\end{equation}
may cause the estimate to converge to the wrong value. In these cases EFC alone cannot prevent the rise in the intensity of the speckles. A similar failure of the ``zero probe''
EKF has been reported in the context of dark hole creation (\cite{riggs2014optimal}).

Consequently, we propose adding some small control perturbation (dither)
$\delta\mathbf{u}$ at every time step,

\begin{equation}
\Delta\mathbf{u}(k)=\Delta\mathbf{u}_{opt}(k)+\delta\mathbf{u}(k),\label{eq:EFC_with_dither}
\end{equation}
in order to introduce phase diversity and avoid the divergence of the
EKF. For simplicity, we suggest randomly choosing

\begin{equation}
\delta\mathbf{u}(k)\sim{\cal N}\left(0,\sigma_{u}^{2}{\cal I}\right),
\end{equation}
where $\sigma_{u}$ is small to ensure that the average intensity
in the dark hole is not significantly increased. An empirical approach
to choosing $\sigma_{u}$ is presented in section \ref{sub:open_loop_simulation}.

\subsection{\label{sub:recalibration}The Effects of Errors in $\cal G$ and Dark Hole Recalibration}

While errors in the control interaction matrix, the Jacobian $\cal G$, affect both the real time and the post-processing estimates, the authors found these effects to not be very significant. Specifically, one of the implicit benefits of the above closed loop approach is that it naturally compensates for the increase in speckles intensity caused by errors in the estimate. Moreover, the EFC controller has a regularization component which counteracts zero mean errors in $G_{ij}$ (assuming they are uncorrelated between pixels). We also suspect that the random choice of $\delta\mathbf{u}$ might have a similar regularization effect in post-processing, although we leave a detailed robustness analysis for future work.

The effect of model errors can, however, be noticed after long integration times (see Fig.~\ref{fig:closed_loop_EKF}
in section \ref{sub:closed_loop_simulation}).
As the phase at the entrance pupil drifts, the electric field, $E^{OL}$, its estimate, $\hat{E}^{OL}$,
and the control deviation, $\Delta\mathbf{u}$, all increase; hence,
the linearization in Eq.~(\ref{eq:EOL_def}) becomes less accurate. To
alleviate the accumulation of systematic errors ($G_{ij}$ multiplied
by a large $\Delta\mathbf{u}$), we suggest periodically shifting
$\mathbf{u}_{0}$ and the speckle estimates

\begin{eqnarray}
\mathbf{u}_{0}\left(k_{r}\right) \leftarrow &  \mathbf{u}_{0}\left(k_{r}\right)+\Delta\mathbf{u}\left(k_{r}\right)\label{eq:recalib_u}\\
\hat{E}_{ij}^{OL}\left(k_{r}\right) \leftarrow & \hat{E}_{ij}^{OL}\left(k_{r}\right)+G_{ij}\Delta\mathbf{u}\left(k_{r}\right),\label{eq:recalib_E}
\end{eqnarray}
where ``$\leftarrow$'' denotes assignment and $k_{r}$ is some predetermined number of time steps. In other words, we suggest updating the nominal dark hole DM setting, around which the system is linearized, every $k_{r}$ steps.
Such periodic ``recalibration'' eliminates the systematic bias at the expense
of a temporarily higher estimation error, which is then corrected
at future iterations.

One can think of the resulting closed loop scheme as performing one step of a dark hole creation algorithm for every $k_{r}$ steps of the newly suggested dark hole maintenance algorithm. Indeed, the former uses large actuations to quickly create a dark hole while the latter uses finer control inputs to battle small drifts in the speckle field resulting in better final estimates. 

\section{\label{sec:numerical_results}Numerical Simulations}

In order to make the discussion of the numerical results more general,
we identify several non-dimensional parameters and calculate their
typical values for a mission similar to WFIRST--CGI observing a relatively
faint target planet.

\begin{figure}
\includegraphics[width=\textwidth]{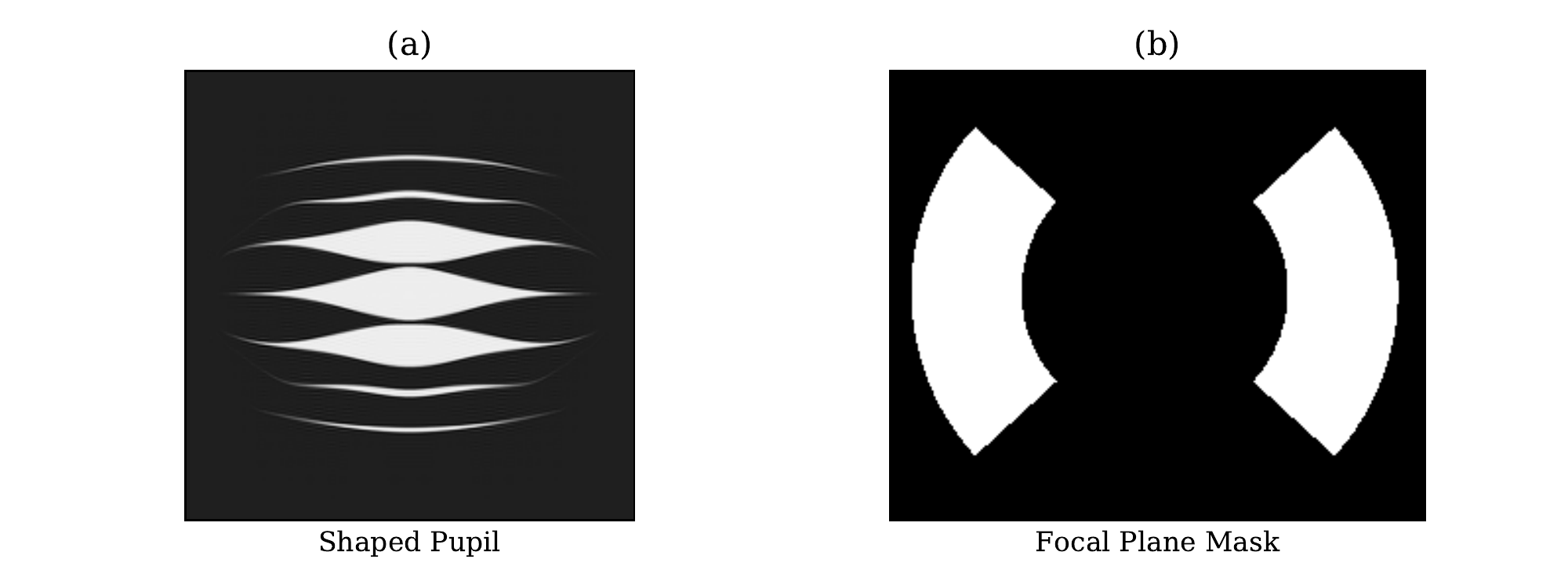}
\caption{\label{fig:masks}The shape pupil (a) introduced in \cite{belikov2007demonstration} and used together with the focal plane mask (b) in our simulation of the HCIL testbed.}
\end{figure}

Consider a $2.37$ meter telescope observing 55 Cancri (a magnitude 6 star) in
$10\%$ broadband light centered at $635\:\mathrm{nm}$. We simulated
the Princeton High Contrast Imaging Lab (HCIL) testbed with a shaped pupil coronagraph (described in detail in \cite{belikov2007demonstration} and shown in Fig.~\ref{fig:masks}(a))
with a contrast of $3.2\cdot10^{-9}$ at $8\:\lambda/D$ where the target planet, 55 Cancri d, would be located. The detector mask (Fig.~\ref{fig:masks}(b)) consists of $2088$ pixels spanning two $[-45^\circ,45^\circ]\times[6.5\:\lambda/D,11\:\lambda/D]$ slices. While the shaped pupil used here is unlike that being planned for WFIRST, it is representative of a shaped pupil coronagraph and allows easy comparison to lab results.  Future work will explicitly model the WFIRST/CGI optical system. 

When taking optical losses into account, the above setup results in an average of
$5\cdot10^{-3}$ photons from the star reaching a detector pixel every second and a peak of $2\cdot10^{-3}$ photons per second from the planet. We assumed all other noise sources (including zodi and dark current) contributed an additional $3\cdot10^{-3}$ photons per second per pixel. Fixing a $100$ second exposure time allows us to define non-dimensional parameters for the scaled
intensities of the star and the planet in terms of the average number
of photons per pixel per frame (see table \ref{tab:parameters}).

\begin{table}[h!]
\begin{tabular}{lcc}
Parameter & Expression & Order of Magnitude \\
\hline

Mean speckles intensity & $\frac{1}{\left|P\right|}\underset{i,j\in P}{\sum}\left|E_{ij}(0)\right|^{2}$ & $1\frac{\#photons}{pixel\cdot frame}$ \\
Typical planet intensity & $\frac{1}{2}\underset{i,j}{\max}I_{ij}^{P}$ & $0.1\frac{\#photons}{pixel\cdot frame}$ \\
Dark current effective intensity & $I^{D}$ & $0.1\frac{\#photons}{pixel\cdot frame}$ \\
Field drift variance & $\left\Vert \Sigma_{ij}\right\Vert $ & $0.01-0.1\frac{\#photons}{pixel\cdot\left(frame\right)^{2}}$
\end{tabular}
\caption{\label{tab:parameters}Non-dimensional parameters that determine the
intensity ratios between the star, the planet, the dark current and
speckle instabilities. The low number of photons suggests that a Gaussian
approximation in section \ref{sub:high_intensity_estimation} might
be applicable to the speckles, but not to the incoherent light.}

\end{table}

Finally, to characterize the speckle drift at the focal plane, we introduced
phase perturbations at the pupil plane and propagated them through
the optical system to approximate the statistics of the electric field prior, $\mu_{ij},\Sigma_{ij},\mu_{ij,0},\Sigma_{ij,0}$
(see Eqs. (\ref{eq:field_increment}),(\ref{eq:field_init})).
Although our estimator doesn't rely on any particular phase disturbance
model, for intuitive purposes we chose to represent them using the  first 15 Zernike polynomials whose coefficients satisfy a random walk with amplitude $0.1\:\mathrm{nm}$
r.m.s. (per frame) (a relatively
unstable system according to \cite{shaklan2011stability}). Propagating
this or any other type of disturbances gives the fourth non-dimensional
parameter in table \ref{tab:parameters}.

In the remainder of this section, we explore the performance of the offline estimator
and the online controller in this setting. The offline procedure was implemented as a standalone algorithm, described in Sec.~\ref{sub:implementation}, to compute an estimate of the incoherent intensity given photon measurements and DM controls, whether they were simulated in an open loop (Sec.~\ref{sub:open_loop_simulation}) or closed loop (Sec.~\ref{sub:closed_loop_simulation}) manner. The online controller (described in Appendix \ref{sec:ekf_equations} and employed in Sec.~\ref{sub:closed_loop_simulation}), was implemented as a part of our optics model and became active immediately after the dark hole creation step. The measurements, $y_{ij}(k)$, were sampled from a Poisson distribution corresponding to the simulated speckle field and incoherent intensity at pixel $i,j$ and frame $k$.

\subsection{\label{sub:open_loop_simulation}Offline Estimator with Small Drift}

To evaluate the offline intensity estimator in an open loop scenario
we performed a Monte Carlo study in which the sample path of the electric
field and the incoherent intensity were estimated for each realization
of the drift process. 

All simulations considered a single pixel (the indices $ij$ are dropped)
at which the incoherent intensity remained constant at $I=0.22$.
Each path of the real and imaginary parts of the electric field was
a realization of a Brownian bridge between frames $k=0$
and $k=T$, that is, the increments of the real and imaginary parts of the
field were normally distributed,

\begin{equation}
\mathrm{Re}\left\{ E^{OL}(k+1)-E^{OL}(k)\right\} \sim{\cal N}\left(0,\sigma^{2}\right)
\end{equation}
where $\sigma=0.03$ and $E^{OL}(0)=E^{OL}(T)=0$. This last constraint
resulted in a ``small'' drift so that we could choose a fixed $\mathbf{u}_{opt}=0$
and focus only on the effects of the dither magnitude. To this end,
the real and imaginary parts of the actuator term were also normally
distributed,

\begin{equation}
\mathrm{Re}\left\{ G\Delta\mathbf{u}\right\} \sim{\cal N}\left(0,\sigma_{u}^{2}\right)
\end{equation}
where the sensitivity to the dither magnitude $\sigma_{u}$ (in $\sqrt{\frac{\#photons}{pixel\cdot frame}}$),
was the subject of investigation in this section.

\begin{figure}
\includegraphics[width=\textwidth]{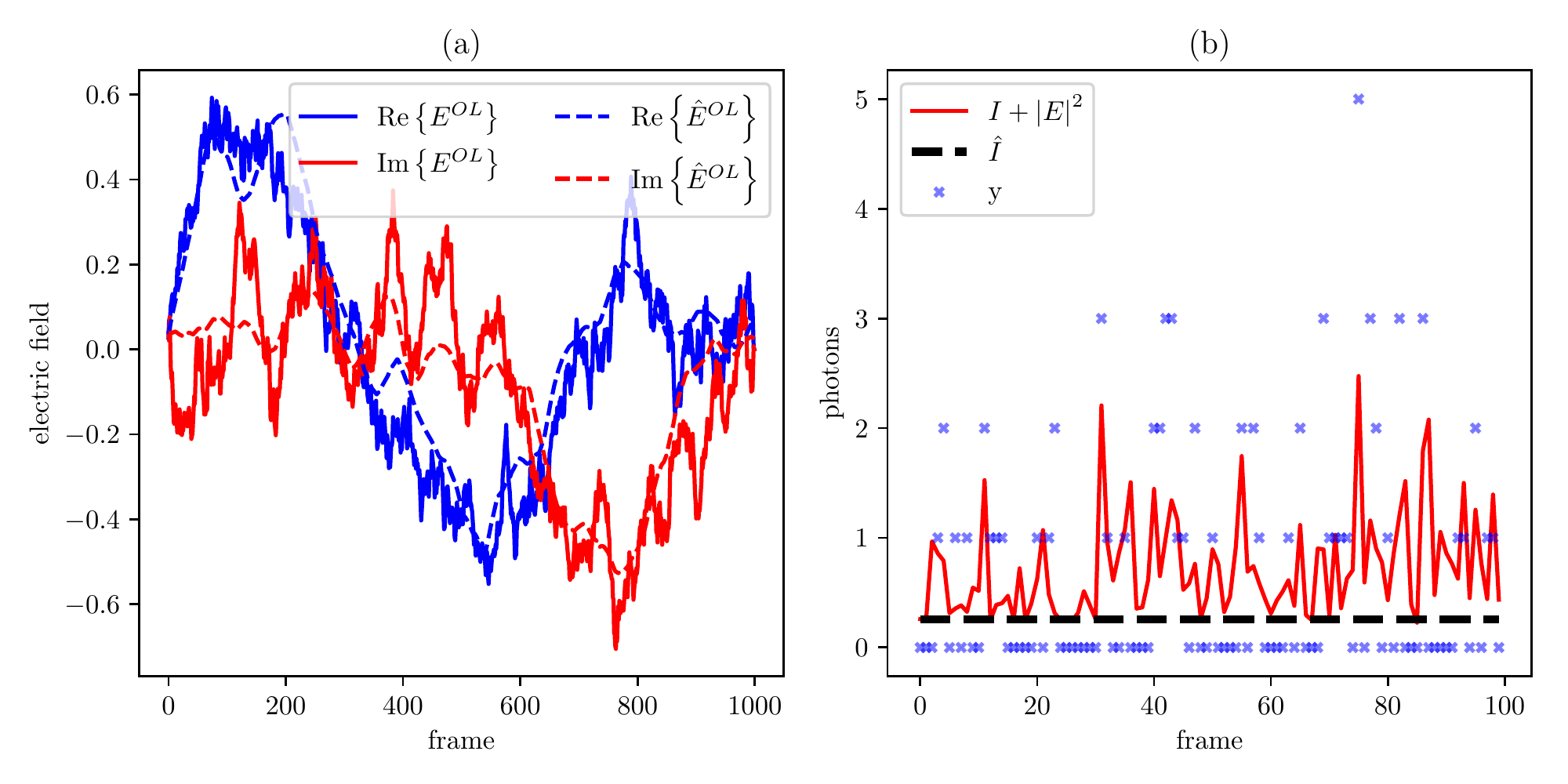}
\caption{\label{fig:offline_process}(a) A sample path of the Brownian bridge of the simulated open loop electric field (solid lines) and its estimate (dashed lines). (b) Total light intensity including the effect of the dither
with $\sigma_{u}=0.5$ (solid line), measured number of photons ($\cdot$)
and the incoherent light estimate (dashed line). Only the first 100 frames are shown.}
\end{figure}

A sample path of the drift process is shown in Fig.~\ref{fig:offline_process}(a)
while Fig.~\ref{fig:offline_process}(b) shows the number of photons
``detected'' at each frame. The estimates of the speckle sample
path (assuming $E(0)=0)$ and the incoherent light, both obtained
via the algorithm described in section \ref{sub:estimation_algorithm},
are also shown for comparison. 

\begin{figure}
\includegraphics[width=\textwidth]{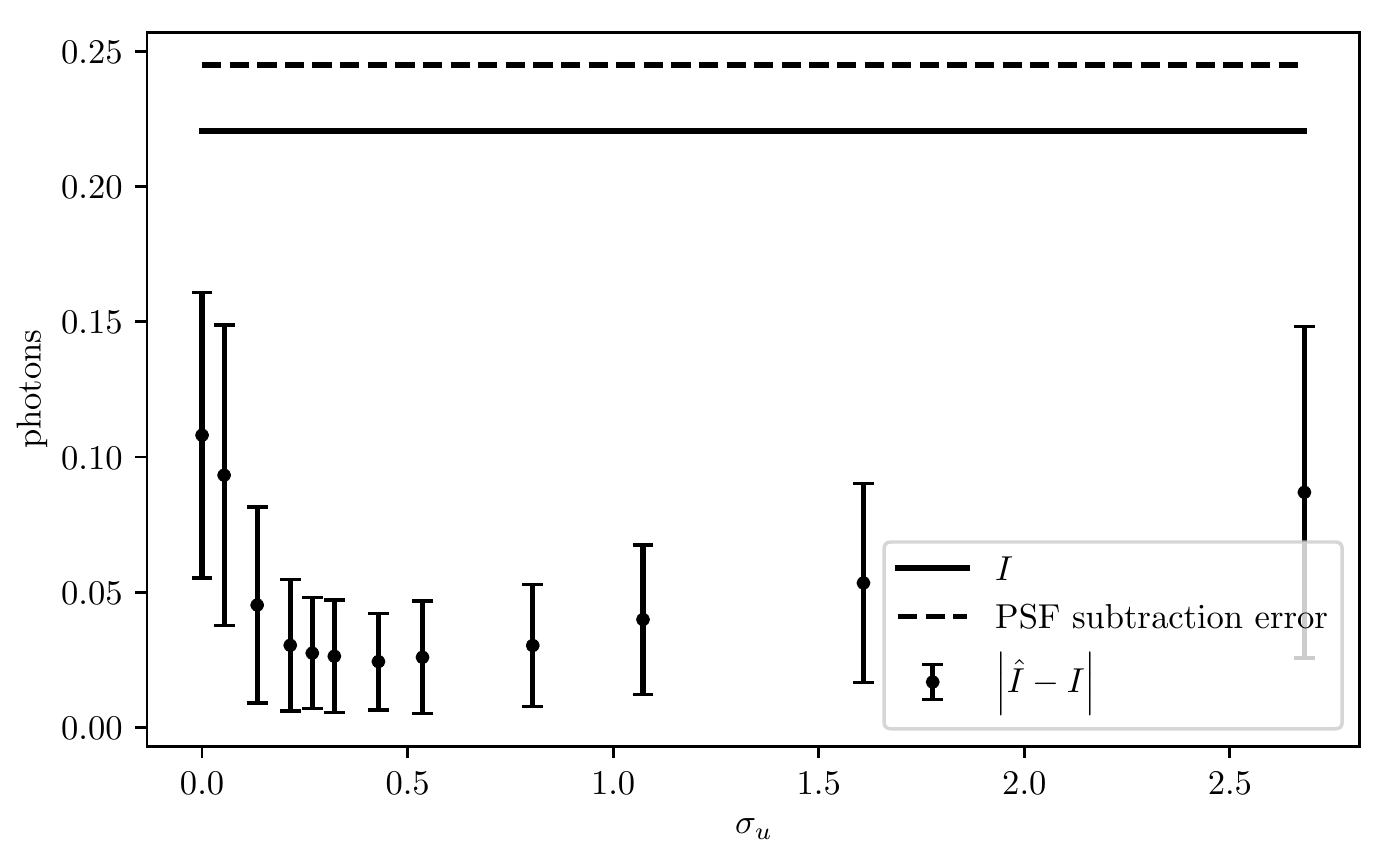}
\caption{\label{fig:offline_errors}Comparison of errors in intensity (mean
and standard deviation) as a function of dither intensity $\sigma_{u}$
across numerous Monte-Carlo simulations. Higher dither intensities
increase the number of photons (in figure \ref{fig:offline_process}(b))
and the associated shot noise but also the phase diversity of the
signal. PSF subtraction (dashed line) does not take drift into account
and gives very large errors. The incoherent light intensity is shown
for comparison (solid line) with no dark current ($I_{D}=0$) present in
this simulation.}
\end{figure}

In order to assess the effect and necessity of the dither, we computed
the average error in the estimate of the incoherent intensity, $I$, across a Monte Carlo ensemble with several
dither magnitudes, $\sigma_{u}$. As can be seen in Fig.~\ref{fig:offline_errors},
even without dithering ($\sigma_{u}=0$), the estimation algorithm
significantly outperforms PSF subtraction. For small non-zero values
of $\sigma_{u}$, the errors become smaller and less spread
out (an indicator of a smaller variance of the estimator). However,
larger dither lets more star light into the dark hole and the associated
shot noise drives the error back up. We conclude that there exists
an optimal non-zero value of $\sigma_{u}$ that reduces the mean
error and variance of the proposed estimator. Estimating the precise optimal $\sigma_{u}$ for a real system such as WFIRST/CGI requires the capacity to simulate numerous observation scenarios of the system (\cite{riggs2018fast}) with meaningful values of the WFE drift (\cite{seo2017hybrid}).
Nevertheless, Fig.~\ref{fig:offline_errors} suggests that there is a wide range of near-optimum dither amplitudes in which the estimate errors remain almost constant.

\subsection{\label{sub:closed_loop_simulation}Closed Loop with Unbounded Drift}

\begin{figure}
\includegraphics[width=\textwidth]{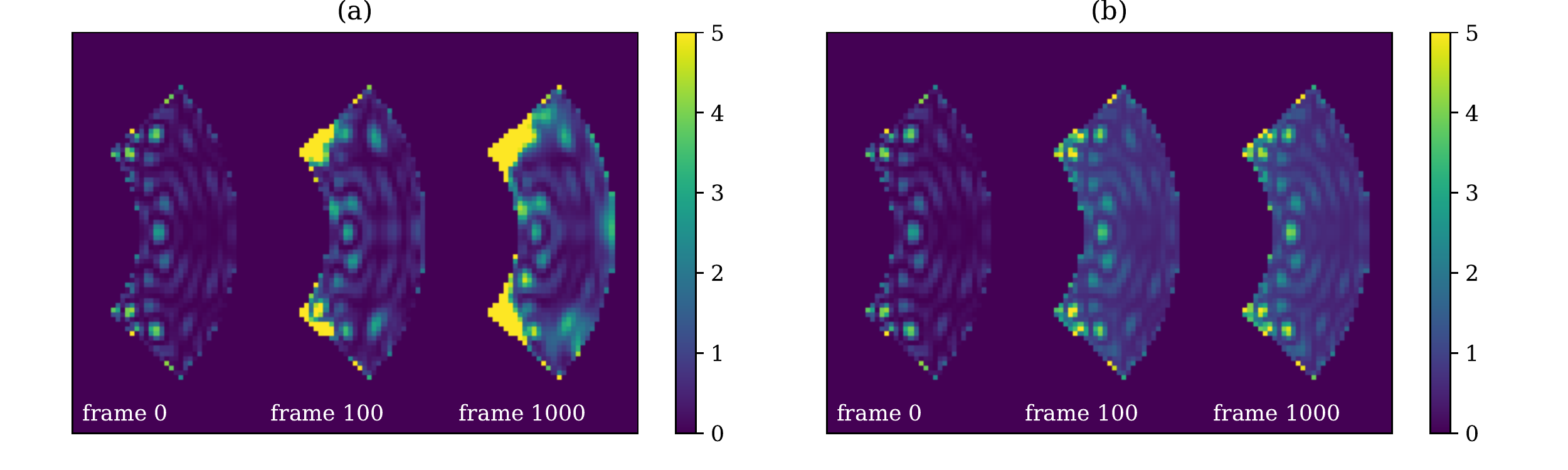}
\caption{\label{fig:speckles_image}(a) Speckle intensity increases without
bound as the WFE performs a random walk (in terms of the first 15
Zernike polynomials). (b) Closing the loop on the electric field helps
maintain a high contrast despite the drift. The contrast is slightly
worse than in a perfect dark hole due to dithering.}
\end{figure}

The maximum expected WFE is commonly used to estimate the best attainable SNR
or to specify stability requirements of the optical system (\cite{krist2008extraction,nemati2017sensitivity,shaklan2011stability}).
In this section, however, we assume that the pupil plane phase perturbations
perform an unbounded random walk. In that case, the intensities of the speckles in the
dark hole also grow without bounds, as shown in Fig.~\ref{fig:speckles_image}(a).

Maintaining a high contrast in the dark hole is achieved through a
combination of the EKF and EFC described in Sections \ref{sub:high_intensity_estimation}
and \ref{sub:EFC_dithering}. The resulting speckle field is shown
in Fig.~\ref{fig:speckles_image}(b) and although the intensity
of the speckles is higher than in the perfect initial dark hole, their magnitude remains bounded.

\begin{figure}
\includegraphics[width=\textwidth]{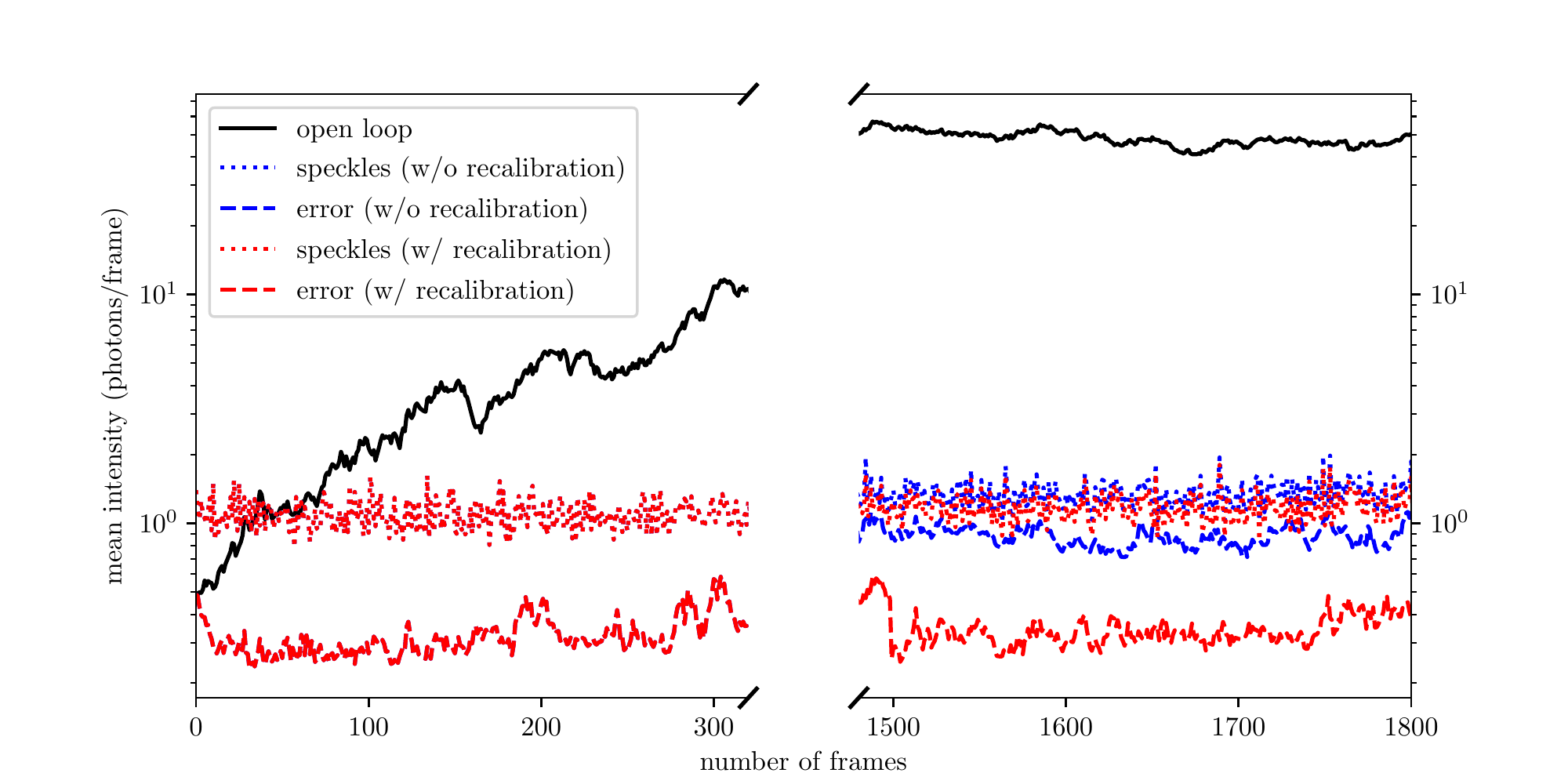}
\caption{\label{fig:closed_loop_EKF}In open loop, as the phase errors at the pupil plane drift, the average intensity of the speckles steadily increases (solid
line). However, the approach outlined in section \ref{sec:closed_loop}
maintains the intensity of the speckles at a constant level (dashed line).
As the drift increases, the accumulated systematic error due to the
linearization (Eq.~(\ref{eq:EOL_def})) becomes significant
(blue dotted line). This can be remedied by periodically recalibrating
the dark hole control setting as discussed in section \ref{sub:recalibration}
(red dotted line).}
\end{figure}

Looking at the mean intensities and errors in Fig.~\ref{fig:closed_loop_EKF},
the open loop dark hole begins at a high contrast which quickly deteriorates,
while the dark hole maintenance controller keeps the contrast at a constant level. This level is slightly worse than the initial contrast due to the dither ($\sigma_{u}\ne0$) which was shown to be necessary for
accurate estimation in section \ref{sub:open_loop_simulation}. Note
that we allowed a relatively large drift, hence PSF subtraction would
perform poorly in this case: at earlier times a low number of photons
would result in low SNR, and at later times the error in the PSF would
dominate.

The necessity of recalibrating the dark hole control is also apparent
from Fig.~\ref{fig:closed_loop_EKF}. With constant $\mathbf{u}_{0}$,
the linearization in Eq.~(\ref{eq:EOL_def}) becomes inaccurate as the
open loop field drifts by a considerable amount. However, a simple
recalibration step (Eqs. (\ref{eq:recalib_u}),(\ref{eq:recalib_E}))
every 500 frames resolves the issue.

\begin{figure}
\includegraphics[scale=0.7]{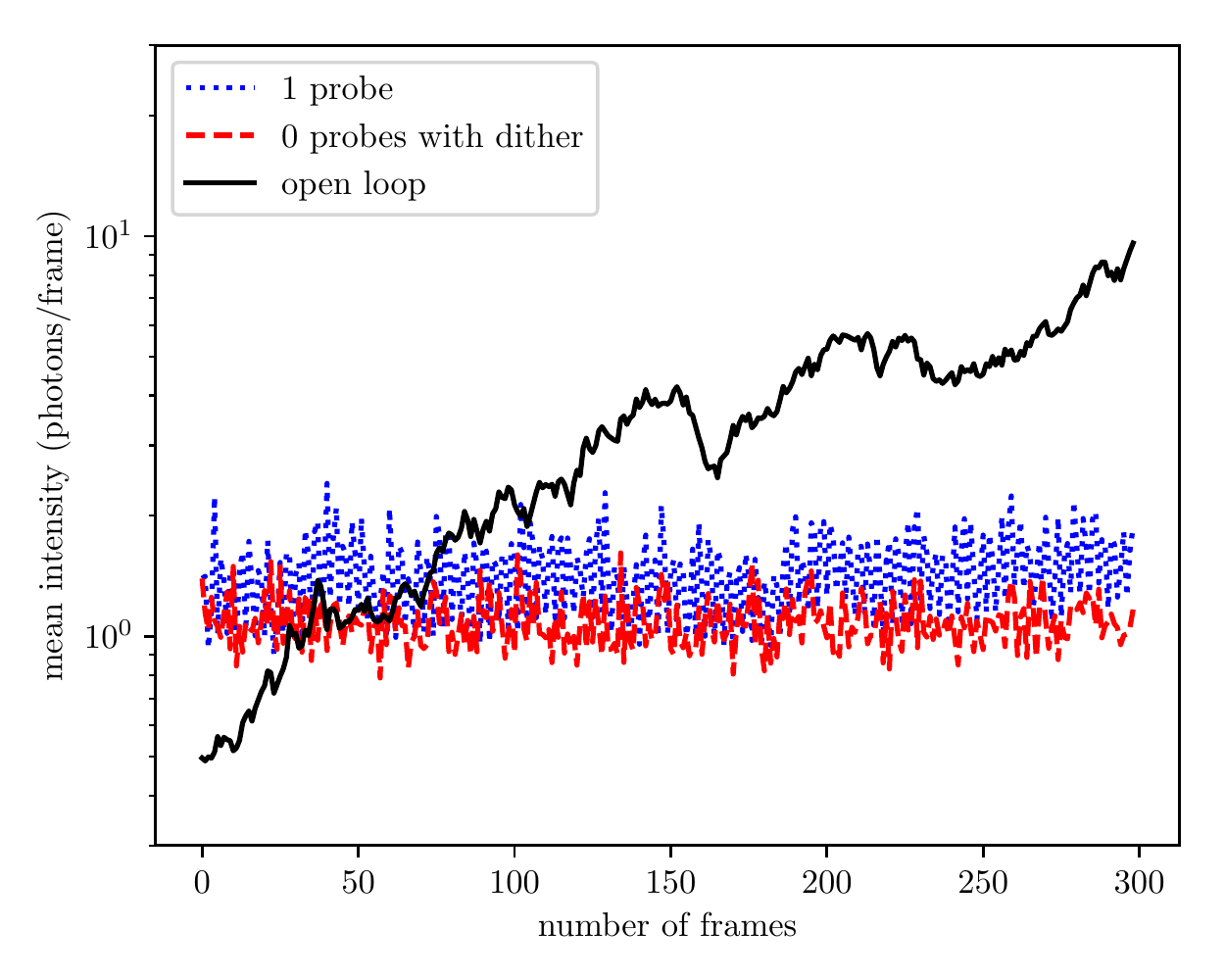}
\centering
\caption{\label{fig:closed_loop_comparison}The newly suggested online control law
(red dashed line) slightly outperforms the 1-probe EKF 
(blue dotted line) which was designed for creating the dark hole and
doesn't take the WFE drift into account (\cite{groff2013kalman}). The intensity refers to speckles only; its effect on the post-processing estimate of the incoherent intensity are shown in Fig.~\ref{fig:combined}.}
\end{figure}

Finally, the newly suggested EKF with dithering is simpler,
computationally more efficient and gives more accurate results than the
previously suggested 1-probe EKF algorithm (\cite{groff2013kalman}).
We attribute this to the fact that the latter was designed for creating
the dark hole when speckle intensities are high but remain constant
over time. Figure. \ref{fig:closed_loop_comparison} shows the dark hole intensity at each of the three frames of the 1-probe EKF: one frame with an ``unperturbed'' control and two frames with complex-conjugate perturbations. We observe that during all three frames, including those corresponding to the ``unperturbed'' control, the contrast in the dark hole was comparable or worse than the contrast maintained by the newly suggested algorithm. Using larger probes would result in better estimates for the 1-probe EKF and hence better contrast during one third of the duty cycle, but would also dramatically increase the intensity of the speckles during the other two thirds. We also note that the EKF version described in detail in Appendix \ref{sec:ekf_equations} is more precise than the version in \cite{groff2013kalman} (since it takes drift into account), regardless of the method chosen to introduce phase diversity (probes or dither).

\subsection{Combined Results}

Both dark hole maintenance (Section \ref{sec:closed_loop}) and post-processing (Section \ref{sec:a_posterior_estimation}) are necessary
for accurate estimation of the incoherent light. Although the recursive
estimate of the incoherent light from the EKF is very noisy, it is sufficient to
keep the speckles bounded. Using only classical PSF subtraction, even if the measurements are acquired from a ``closed-loop-maintained'' dark hole, yields poor results when the incoherent light is dimmer than the time varying speckles. The offline estimator was specifically designed to address this issue. However, when used on measurements from an ``un-maintained'' dark hole, it suffers from progressively higher errors as the open loop shot noise increases due to the speckle drift.

\begin{figure}
\includegraphics[width=\textwidth]{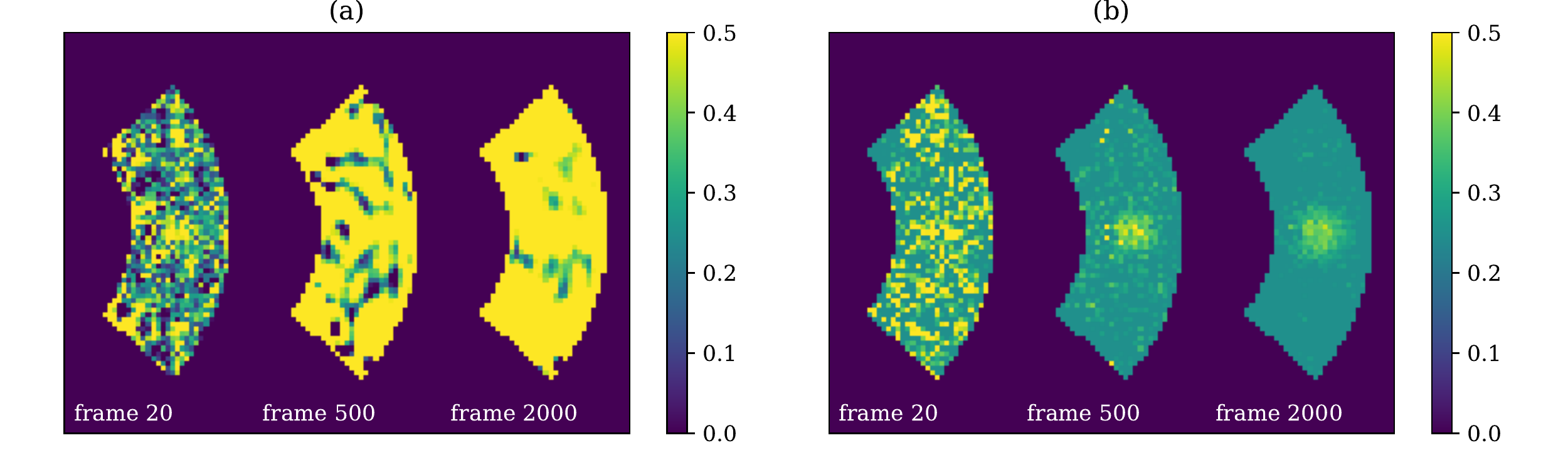}
\caption{\label{fig:incoherent_image}(a) Incoherent intensity estimation via classical
PSF subtraction is completely ruined by the open loop speckle drift when the DMs are kept fixed.
(b) Closed loop dark hole maintenance followed by an offline {\it a posteriori}
intensity estimation, allows a significantly larger SNR to be achieved given
the same integration times. Note that the incoherent intensity includes the dark current.}
\end{figure}

As can be seen in Fig.~\ref{fig:incoherent_image}(a), the effect
of unbounded WFE drift when using PSF subtraction is highly detrimental. In comparison,
the combined control and estimation approach maintains the dark hole
for as long as necessary 
and the incoherent light estimate becomes
progressively better (Fig.~\ref{fig:incoherent_image}(b)). The
effective dark current intensity, which we considered to be part of
the total incoherent intensity, is known, can be subtracted, and therefore
doesn't limit the SNR. However, as the magnitude of the drift increases, so does the optimal magnitude of the dither ($\sigma_{u}$) and the shot noise associated with it. This means that for extremely large dither, extremely long integration times would be necessary and the linearity assumption (Eq.~\ref{eq:linearity_assumption}) would break down, effectively limiting the achievable SNR.

\begin{figure}
\includegraphics[width=\textwidth]{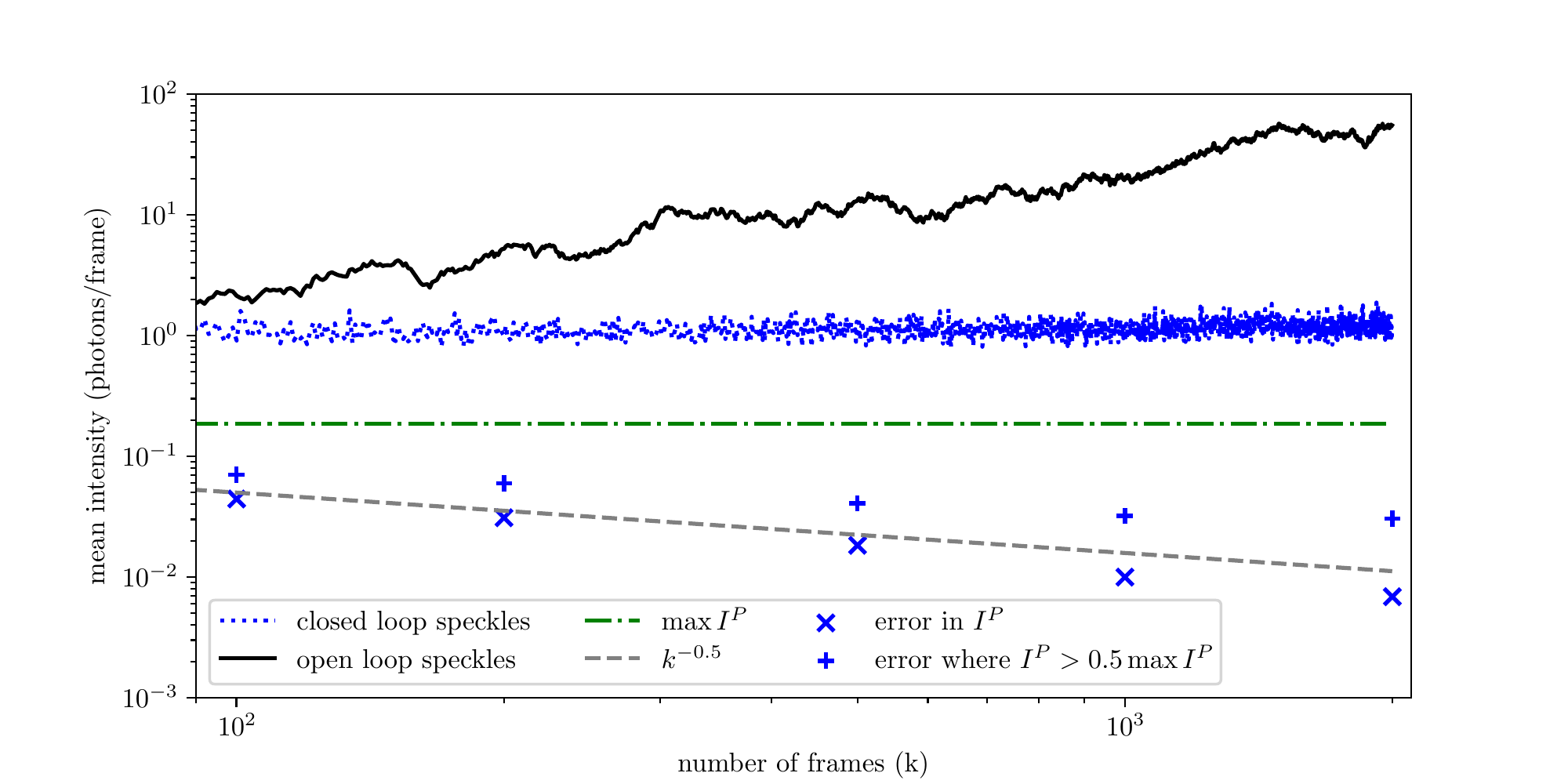}
\caption{\label{fig:combined}Mean intensity of open loop (solid black line) and
closed loop (dotted blue line) evolution of the speckles. In our simulation,
the planet was about five times dimmer than the closed loop speckles requiring a post-processing
step to estimate its intensity (dash-dotted line). 
The errors in the
intensity (computed offline and denoted as $\times$ and $+$), decay
roughly as the square root of the integration time (dashed line).}
\end{figure}

In a perfect linear model of the observatory, this approach eliminates the systematic PSF error which,
according to Eq.~(\ref{eq:SNR}), means that  the incoherent
intensity error should decrease as the square root of the number of
frames (that is, the estimation becomes photon noise limited). This was nearly the case in our simulation  illustrated
in Fig.~\ref{fig:combined}, where the error in the half-max planet intensity region ($I^{P}>0.5\max I^{P}$) initially decreases as $k^{-0.5}$ but, after an equivalent of 56 hours of observation on a system like WFIRST, reaches a steady value of  $1/30$ of the mean speckle intensity. 
We suspect that this eventual SNR limitation is due to imperfect estimates of the Jacobian (${\cal G}$, which varies with time due to the drift) and the error statistics ($\Sigma_{ij},\mu_{ij}$).

\section{Conclusions}

In this paper, we introduced two algorithms for focal plane wavefront control
and estimation of speckle intensity evolution and incoherent intensities. Combined,
these two algorithms maintain the dark hole during long integration
times on the target star and reduce systematic errors present in
PSF subtraction methods. 
Under the admittedly limiting assumptions of a perfectly accurate and linear model of speckle drift and control actuators, we have thus reduced the estimation problem from being background limited to photon noise limited regardless of the flux ratio between the star and planet, the size of the residual speckles, or their time variation. This implies that one can eliminate any bounds on the SNR and therefore detect arbitrarily dim planet given sufficiently long integration times when the two algorithms operate in perfect conditions.

The EKF based algorithm for closed loop dark hole maintenance employs the EFC control law while adding  small DM dither.
This addition was justified analytically in Section \ref{sub:EFC_dithering}
and numerically in Section \ref{sub:open_loop_simulation}, although
we leave the optimization of dither shape and magnitude for future work. We found that the newly suggested approach results in lower residual speckle than the EKF
based algorithms for creating the dark hole, as those do not take WFE drift
into account. 

Allowing for DM actuation and possible speckle drift during the
integration phase requires a more nuanced approach for computing
the intensity of sources incoherent with the star light. An {\it a posteriori}
intensity estimation algorithm proposed in Section \ref{sub:estimation_algorithm}
considers the statistics of the drift and shot noise over the whole
history of integration and gives a (sub-optimal) estimate of the incoherent
intensity. Despite discarding all spatial information, this estimator
is still computationally expensive and has to be performed offline.

The currently proposed mode of operation for the WFIRST, as well as for the proposed HabEx mission, involves a pointing sequence alternating between open loop observation of the target and a periodic ``re-creation'' of the dark hole using a reference star (\cite{rizzo2018wfirst}). This approach has the potential to introduce additional mechanically induced wavefront errors during slews and doesn't take the statistics of the drift at the post-processing step into account.  It  therefore results in significant over-specification of the stability of the instrument. Our method eliminates the need for alternating pointing and achieves considerably better performance than PSF subtraction, thus reducing the stability requirements or observation times. 
We leave for a future work the task of estimating these bounds on the wavefront stability for various realistic values of flux ratio between the speckles and the planet, and the sensitivity analysis of the proposed algorithms. This analysis will address model errors stemming from the assumptions of linearity and Brownian motion of the WFE, and the associated imperfect estimates of the Jacobians and drift statistics.

Finally, we note that the proposed estimators were developed for a monochromatic optical model
with no regard for correlation between focal plane pixels. They can,
however, be formulated in a reduced order setting similar to KLIP (\cite{soummer2012detection})
with a particular choice of PSF basis (spatial or chromatic) for the speckle field.
This direction will also be explored in future work.

\section{Acknowledgements}
This work was partially performed under contract to the Jet Propulsion Laboratory of the California Institute of Technology, award number AWD1004079.  We would also like to thank Bijan Nemati for providing technical details for various observation scenarios of the WFIRST telescope.

\appendix

\section{\label{sec:ekf_equations}Dark Hole Maintenance Algorithm}

We provide here the implementation details for the scheme for real-time estimation and correction of the speckle field in the dark hole (Sec.~\ref{sec:closed_loop}). The algorithm is split below into an estimator and a controller, and does not directly rely on the post-processing scheme (Sec.~\ref{sec:a_posterior_estimation}, although the final results benefit from using the two methods in conjunction).

The controller is designed to start operating after a dark hole was created with a nominal DM setting, $\mathbf{u}_{0}$. The control input during frame $k$, $\mathbf{u}(k) = \mathbf{u}_{0} + \Delta \mathbf{u}(k)$, relies on an estimate of the speckles from the previous frame, $\hat{E}_{ij}^{OL}(k-1)$, at each pixel $ij$. When, stacking the real and imaginary parts of all $\hat{E}_{ij}^{OL}(k-1)$ into a column vector $ \hat{\mathbf{E}}^{OL}(k-1)$, the control deviation is given by Eqs. (\ref{eq:opt_u}) and (\ref{eq:EFC_with_dither}),
\begin{equation}
\Delta\mathbf{u}(k)=-\left({\cal G}^{t}{\cal G}+\alpha{\cal I}\right)^{-1}{\cal G}^{t}  \hat{\mathbf{E}}^{OL}(k-1) + \delta\mathbf{u}(k)
\end{equation}
where $\delta\mathbf{u}(k)$ are sampled from ${\cal N}\left(0,\sigma_{u}^{2}{\cal I}\right)$, $\cal G$ is the control Jacobian (its rows consist of real and imaginary parts of all $G_{ij}$), $\cal I$ is the identity matrix and $\alpha, \sigma_{u}$ are some empirically chosen parameters. 

The estimator of the speckle field is an extended Kalman filter (see, for example \cite{stengel1994optimal}) corresponding to the state equations (\ref{eq:EKF_state1}),(\ref{eq:EKF_state2}), the measurement equation (\ref{eq:EKF_observation}) and the process noise  statistics given by Eqs. (\ref{eq:field_increment}), (\ref{eq:field_init}). For simplicity we assume zero mean drifts and prior ($\mu_{ij}=\mu_{ij,0}=0$) and initialize the speckles estimates with $\hat{E}_{ij}^{OL}(0)=0$ and the incoherent light estimates with $\hat{I}_{ij}(0)=I^{D}$ (the effective dark current intensity). Note that the estimator in fact consists of multiple independent Extended Kalman filters, one for each pixel ($i,j$).

The state estimates are advanced via
\begin{equation}
\begin{bmatrix}\mathrm{Re}\left\{ \hat{E}_{ij}^{OL}(k+1)\right\} \\
\mathrm{Im}\left\{ \hat{E}_{ij}^{OL}(k+1)\right\} \\
\hat{I}_{ij}(k+1)
\end{bmatrix}=	\begin{bmatrix}\mathrm{Re}\left\{ \hat{E}_{ij}^{OL}(k)\right\} \\
\mathrm{Im}\left\{ \hat{E}_{ij}^{OL}(k)\right\} \\
\hat{I}_{ij}(k)
\end{bmatrix}+K_{ij}(k)\left(y_{ij}(k)-\hat{y}_{ij}(k)\right)
\end{equation}
where $y_{ij}(k)$ is the number of photons detected at pixel $i,j$ during frame $k$,
\begin{equation}
\hat{y}_{ij}(k)=\max\left\{\hat{I}_{ij}(k),I^{D}\right\}+\left|\hat{E}_{ij}^{OL}(k)+G_{ij}\Delta\mathbf{u}(k)\right|^{2}
\end{equation}
and $K_{ij}(k)$ is the Kalman gain. The gain is given by
\begin{equation}
 K_{ij}(k)=\frac{1}{H_{ij}(k)P_{ij}(k|k-1)H_{ij}(k)^{T}+\hat{y}_{ij}(k) }P_{ij}(k|k-1)H_{ij}(k)^{T},
\end{equation}
where
\begin{equation}
H_{ij}(k)=\begin{bmatrix}2\mathrm{Re}\left\{ E_{ij}^{OL}(k)+G_{ij}\Delta\mathbf{u}(k)\right\}  & 2\mathrm{Im}\left\{ E_{ij}^{OL}(k)+G_{ij}\Delta\mathbf{u}(k)\right\}  & 1\end{bmatrix},
\end{equation}
and the update-step covariance matrix is approximated via
\begin{equation}
P_{ij}(k+1|k)=\left(I-K_{ij}(k)H_{ij}(k)\right)P_{ij}(k|k-1)+\begin{bmatrix}\Sigma_{ij} & 0\\
0 & \sigma_{I}^{2}
\end{bmatrix}
\end{equation}
and initialized by
\begin{equation}
P_{ij}(1|0)=\begin{bmatrix}\Sigma_{ij,0}+\Sigma_{ij} & 0\\
0 & \sigma_{I,0}^{2}+\sigma_{I}^{2}.
\end{bmatrix}
\end{equation}
In a numerical simulation setting, the parameters $\Sigma_{ij}$, $\Sigma_{ij,0}$ can be approximated by the empirical covariances of the simulated electric field drift and the initial electric field error at pixel $i,j$ respectively. The quantities $\sigma_{I,0}^{2}$ and $\sigma_{I}^{2}$, on the other hand, are somewhat artificial and are subject to fine tuning.

Note that since the controller doesn't utilize the incoherent intensity estimate, the above filter can be simplified by choosing a constant $\hat{I}_{ij}(k)=I^{D}$ in the speckle dominated regimes.

Finally, if the performance of the above algorithm deteriorates for large $k$, one may perform a recalibration every $k_{r}$ steps, as follows (Eqs. (\ref{eq:recalib_u}) and (\ref{eq:recalib_E})):
\begin{eqnarray}
\mathbf{u}_{0}\left(k_{r}\right) \leftarrow & \mathbf{u}_{0}\left(k_{r}\right)+\Delta\mathbf{u}\left(k_{r}\right)\\
\hat{E}_{ij}^{OL}\left(k_{r}\right) \leftarrow & \hat{E}_{ij}^{OL}\left(k_{r}\right)+G_{ij}\Delta\mathbf{u}\left(k_{r}\right).
\end{eqnarray}

\bibliography{DarkHoleDither}
\bibliographystyle{plain}

\end{document}